\documentclass{article}
\usepackage[displaymath]{lineno}
\usepackage{icrctc07}
\title{The UHECR spectrum measured at the Pierre Auger Observatory and its astrophysical implications}
\shorttitle{Auger Spectrum and Interpretation}

\authors{T.Yamamoto$^{1}$ for the Pierre Auger Collaboration$^{2}$ }
\shortauthors{Pierre Auger Collaboration}
\afiliations{$^1$ KICP, Enrico Fermi Institute, University of Chicago, Chicago IL USA\\
$^2$ Pierre Auger Observatory, Av  San Mart{\'\i}n Norte 304,(5613) Malarg\"ue, Argentina}
\email{yamamoto@oddjob.uchicago.edu}

\abstract{The Southern part of the Pierre Auger Observatory is nearing
completion, and has been in stable operation since January 2004 while it has
grown in size. The large sample of data collected so far has led to a
significant improvement in the measurement of the energy spectrum of
UHE cosmic rays over
that previously reported by the Pierre Auger Observatory,
both in statistics and in systematic uncertainties.
We summarize two measurements of the energy spectrum, one based on the
high-statistics surface detector data, and the other based on the hybrid data,
where the precision of the fluorescence measurements is enhanced by
additional information from the surface array. The complementarity of
the two approaches is emphasized and results are compared. Possible
astrophysical implications of our measurements, and in particular the
presence of spectral features, are discussed.}

\begin{document}
\maketitle
\nolinenumbers

\section[UHE cosmic ray energy spectrum]{UHE cosmic ray energy spectrum}

The Pierre Auger Observatory measures extensive air showers induced by
 the highest energy events ($E>10^{18}$ eV) using two detection
 techniques.  Firstly, a collection of telescopes is used to measure
 the ultraviolet fluorescence light produced when electrons in the
 shower excite nitrogen molecules in the atmosphere.  This technique
 will be referred as FD (Fluorescence Detector). It measures the
 longitudinal development of the air-shower and can only be used
 during dark and moonless nights, yielding a duty cycle of
 roughly 10\%.  The second technique (called SD for Surface
 Detector) uses an array of water Cherenkov detectors to sample the
 shower front at ground level. The SD has a duty cycle
 of 100\% and the detection efficiency is 100\% for energies above
 10$^{18.5}$ eV (10$^{18.8}$ eV) at zenith angles below (above)
 60$^\circ$.  The showers recorded by the SD are quantified in size
 using the reconstructed signal at 1000 m from the shower axis,
 called S(1000) \cite{Accuracy}. At large zenith angles (above $60^\circ$),
 due to deflection of the shower particles in the geomagnetic
 field, another energy estimator $N_{19}$ is used \cite{Hasrec}. The
 conversion from these two SD estimators to the primary energy could
 be calculated using full Monte Carlo simulations but the lack of
 knowledge of the primary mass and the
 uncertainties in the hadronic models introduce
 large systematics. Therefore we use a subset of showers called \emph{hybrid} events
 that are detected by both the SD and the FD.
 The conversion parameters from the SD estimators to the energy measured by the FD
 then are derived experimentally.
 A comparison of the results of this calibration with the expectations
 from Monte Carlo simulation can be found in \cite{nmu}.
 The FD measures fluorescence light in proportion to the energy
 deposited by the shower, and so the technique is calorimetric. There
 is, however, a small correction to account for the energy deposited in
 the ground by high energy muons and neutrinos. This ``invisible
 energy'' correction has a small dependence on mass and hadronic
 model. The applied correction is based on the average for proton and
 iron showers from the QGSJet model. This correction factor is about
 10\% and its systematic uncertainty contributes 4\% to the total
 uncertainty in FD energy \cite{invisible1, invisible2}.

\begin{figure}
\begin{center}
\includegraphics[width=0.47\textwidth]{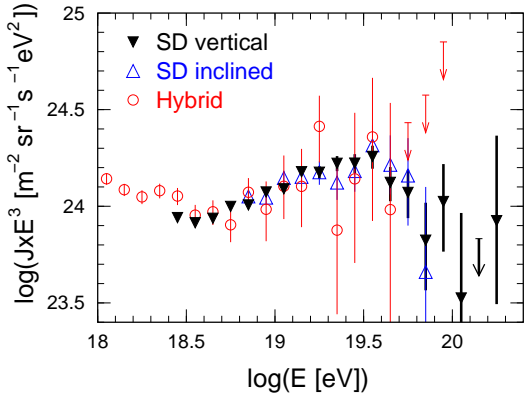}
\end{center}
\caption{The energy spectrum multiplied by $E^3$
 derived from SD using showers at zenith angles above (filled triangles) and
 below (opened triangles) 60$^\circ$
 (\cite{sdspec, Inclinedspec}), together with the spectrum
 derived from the hybrid data set (red circles)\cite{Hybridspec}.
 Arrows indicate 84\% CL upper-limits \cite{web}.
}
\label{fig:augerspec}
\end{figure}

 Fig. \ref{fig:augerspec} shows the energy spectrum multiplied by
 $E^3$ from SD data using showers at zenith angles above and below
 60$^\circ$
 (\cite{sdspec, Inclinedspec}), together with the spectrum derived from the
 $hybrid$ data set (a fluorescence events in coincidence with at least
 one SD station) \cite{Hybridspec}. The agreement between the spectra
 derived using three different
 methods is good and is underpinned by the common method of energy
 calibration based on the FD measurements.
 Therefore all spectra are affected by the 22\% uncertainty
 in the FD energy scale\cite{hybridperform}, in which the largest contribution is
 the absolute fluorescence yield(14\%). In this work
 we have used the fluorescence yield reported in \cite{nagano}.
 This common uncertainty does not affect the relative comparison of our spectra.
 The systematic uncertainty in the hybrid-only spectrum is dominated by
 uncertainties in the calculation of the exposure (16\%).  The
 systematic uncertainty in the SD spectrum has two contributions: the
 calculation of the exposure (3\%) and the statistical uncertainty in
 the calibration of $S(1000)$ and $N_{19}$ with the FD energy
 ($<$10\%). We use a maximum likelihood method, together with our
 knowledge of the systematics, to calculate the relative normalization
 factors necessary to match the spectra with each other. We find that
 the different spectra are in excellent agreement with
 normalization factors smaller than 3\%.
We combine the three spectra weighting each bin based on its statistical
 uncertainty. The final combined spectrum is shown later in
 Fig. \ref{fig:augerspec-model}. 
 It should be noted that the first two bins in the SD spectrum were
 excluded in this procedure. We expect these first two bins are biased by
 threshold effects of the order of 10\%. The deviations of those bins from
 the Hybrid spectrum are in agreement within the systematic uncertainty.

\section[The highest end of the spectrum]{The highest end of the spectrum}

Since the 22\% systematic uncertainty in the energy scale does not
modify the shape of the spectrum, it is possible to check the
continuation of the
spectrum at the highest energies. It could be argued that our energy
calibration has low statistics at the highest energies (see Fig. 3 in
\cite{sdspec}). However no indication of a change in the calibration
parameters with threshold energy used has been found.  A dramatic
change in the hadronic interactions in the energy range where no
hybrid event is observed could also induce false spectral
features. However, there 
is no theoretical basis for such a scenario, and even if it were the
case it will be checked in the future with larger statistics in the
hybrid data set.

\begin{figure}
\begin{center}
\includegraphics[width=0.47\textwidth]{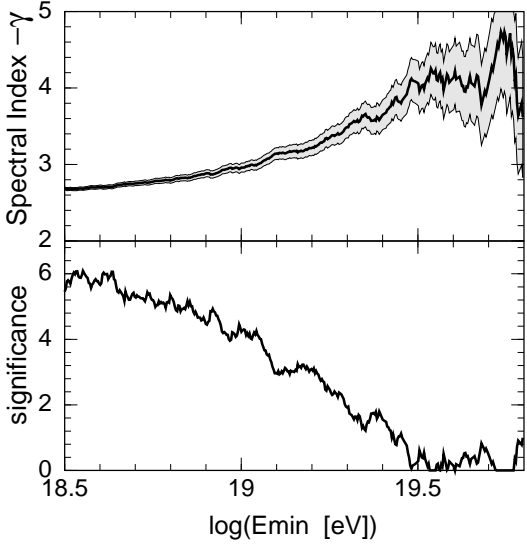}
\end{center}
\caption{Upper panel: Spectral index as a function of minimum energy in
 the fit. Lower panel: significance (in sigma) of the
 deviation from power-law distribution with spectral index from upper panel
 based on the TP statistics.
  }
\label{fig:gamma}
\end{figure}

To check the continuation of the spectrum at the highest energies we
 first fit the SD spectrum between 10$^{18.6}$ eV and 10$^{19.6}$
 eV to a power-law function using a binned likelihood method. The
 spectral index 
 obtained is $\gamma=-2.62\pm$0.03(stat)$\pm$0.02(sys). The
 systematic error is given by the error on the calibration curve in
 \cite{sdspec}. 
 The number of events expected from such a single power-law flux
 above 10$^{19.6}$ eV and 10$^{20}$ eV are
 132$\pm$9 and 30$\pm$2.5 respectively whereas we observe only
 51 events and 2 events. Also, the spectral index from 10$^{19.6}$ eV up to the highest energy
 observed (1.90$\pm$0.16(stat)$\pm$0.20(sys))$\times10^{20}$ eV is
$\gamma=-4.14\pm$0.42(stat) (Fig.\ref{fig:gamma}).
A lack of events at the highest energies is clear.  We then applied
 a statistical test proposed in \cite{tp1}, the so called
 TP-test. The TP statistic
 allows us to test for a power-law distribution on an unbinned
 data set without bias regarding the value of the spectral index.
 Details of this statistical test can be found in \cite{hague}.
%
The upper panel in Fig.\ref{fig:gamma} shows the unbinned maximum-likelihood
estimation of the spectral index ($\gamma$) and its standard deviation
(shaded region) as a function of minimum energy used in the fitting. A clear
change of slope at the highest energy can be seen. The deviation from
the power-law distribution with $\gamma$ shown in this figure is
estimated based on the TP statistic. The lower panel in Fig.\ref{fig:gamma}
shows the estimated deviation in sigma. The hypothesis of the pure
power-law is then rejected with a significance better than 6 sigma and 4
sigma for minimum energies of 10$^{18.6}$ eV and 10$^{19}$ eV
respectively.

\section[Astrophysical interpretation]{Astrophysical interpretation}

\begin{figure}
\begin{center}
\includegraphics[width=0.47\textwidth]{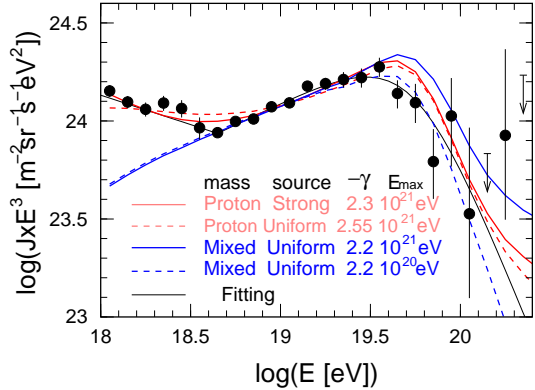}
\end{center}
\caption{The combined energy spectrum multiplied by $E^3$ \cite{web}, together with
 a fit to  Eq. \ref{eq:powfermi} (black line),  and the predictions of
 two astrophysical models (blue and red lines). The input assumptions of
 the models (mass composition at the sources, the source distribution,
 spectral index and exponential cutoff energy per charge at the
 acceleration site) are indicated in the figure.}
\label{fig:augerspec-model}
\end{figure}

In the previous section, we have shown that the rejection of the
 hypothesis of a continuation of the spectrum in the form of a power-law
 is statistically significant. Moreover, a spectral break at
$\sim 10^{18.5}$ eV, the so-called \emph{ankle}, is apparent in
 Fig. \ref{fig:augerspec}. Therefore we
 fitted the combined Auger spectrum to the following equation:
\begin{eqnarray}
J(E;E<E_{ankle})&\propto& E^{\gamma_1} \nonumber \\
J(E;E>E_{ankle})&\propto& E^{\gamma_2} \frac{1}{1+\exp\left(\frac{\mathrm{lg}E-\mathrm{lg}E_c}{W_c}\right)}  \nonumber\\
\label{eq:powfermi}
\end{eqnarray}
where $\gamma_1$ and $\gamma_2$ are the spectral index before and after
the break respectively, $E_{ankle}$ is the position of the break, and
the second term in the second equation is a flux suppression term where
$E_c$ is the energy at which the flux is suppressed 50\% compared to a
pure power-law, and $W_c$ determines the sharpness of the cutoff. Here using a
binned likelihood method, the values of the parameters obtained are
the following: $\gamma_1=-3.30\pm$0.06,
$\gamma_2=-2.56\pm$0.06, $log_{10}E_{ankle}$=18.65$\pm$0.04,
$log_{10}E_c$=19.74$\pm$0.06 and $W_c$=0.16$\pm$0.04.
The $\chi^2/$dof for this fit is 16.7/16.  The
black line in Fig. \ref{fig:augerspec-model} shows the result of the
fit.

Fig. \ref{fig:augerspec-model} shows also a comparison of our data with
some astrophysical models \cite{denis}. 
These models show a flux suppression at the highest energies
(the GZK steepening \cite{gzk1, gzk2}).
The models all assume an injection spectral index, an exponential cutoff
at an energy of $E_{max}$ times the charge of the nucleus, and a mass
composition at the acceleration site as well as a distribution of 
sources. The 
blue lines in the figure assume a mixed composition at the sources,
i.e. with nuclear abundances similar to those of the low-energy galactic
cosmic rays. A uniform distribution of sources and an injection
spectral index of -2.2 (close to the shock acceleration predictions) are
assumed as indicated in the figure. $E_{max}$ is taken as
$10^{20}$ eV (dashed line) and $10^{21}$ eV (solid line). Good agreement
is found down to energies close to $E_{ankle}$.
Below this energy  another component is needed. 

Another set of models which assume only proton primaries
and $E_{max}=10^{21}$ eV are shown by the red lines.
One model assumes uniform source distribution with the spectral index
-2.55 and the other assumes the source evolution has a strong redshift
dependence $(1+z)^5$ with the spectral index -2.3.
It has been suggested that the spectral break at $E_{ankle}$ can be
explained as a feature of the propagation of a pure proton flux in the extragalactic
media including $e^\pm$ pair production \cite{Berezinsky}.
To reproduce our spectrum by this model, we need
a very stronger source evolution. The distribution of the longitudinal
profiles of the showers observed by the FD also disfavors the pure
proton assumption \cite{Xmax}.


\section[Conclusions]{Conclusions}

 Using data from the southern-hemisphere Pierre Auger Observatory, we reject the
 hypothesis that the cosmic ray spectrum continues in the form
 of a power-law above an energy of $10^{19.6}$ eV with 6 sigma significance.
 This result is independent of the systematic uncertainties
 in the energy scale. A precise measurement of the energy spectrum,
 together with anisotropy and mass composition studies in this energy
 range, will shed light on
 the origin of the highest energy particles observed in nature.

\bibliography{libros}

\bibliographystyle{plain}
\end{document}